\documentclass{article}
\usepackage[utf8]{inputenc}
\RequirePackage{amsthm,amsmath}
\RequirePackage{natbib}
\RequirePackage[colorlinks,citecolor=blue,urlcolor=blue]{hyperref}
\RequirePackage[hmarginratio=1:1,top=32mm,columnsep=20pt]{geometry} 
\RequirePackage{textgreek}
\RequirePackage{bm, bbm, blkarray, adjustbox}
\RequirePackage{stackengine}
\RequirePackage{amsfonts}
\RequirePackage{amssymb}
\RequirePackage{arydshln}
\RequirePackage{graphicx} 
\RequirePackage{caption}
\RequirePackage{booktabs}
\RequirePackage{scrextend}
\RequirePackage{ulem}
\usepackage[flushleft]{threeparttable}
\usepackage{lipsum}

\usepackage{caption}
\usepackage{subcaption}

\usepackage{placeins} 
\newcommand{\beginsupplement}{%
    \setcounter{table}{0}
    \renewcommand{\thetable}{S\arabic{table}}%
    \setcounter{figure}{0}
    \renewcommand{\thefigure}{S\arabic{figure}}%
    \renewcommand{\thesection}{S\arabic{section}}%
}

\usepackage{authblk}

\title{PliableBVS: A flexible Bayesian variable selection method for modeling interactions with mandatory modifying variables}

\author[1]{Theophilus Quachie Asenso \thanks{\texttt{CONTACT Theophilus Quachie Asenso. Email: t.q.asenso@medisin.uio.no}}}
\author[1]{Zhi Zhao}
\author[3]{Maren-Helene Langeland Degnes}
\author[4]{Marie Cecilie Paasche Roland}
\author[3,5]{Trond Melbye Michelsen}
\author[1,2]{Manuela Zucknick}
\affil[1]{Oslo Centre for Biostatistics and Epidemiology (OCBE), University of Oslo, Norway}
\affil[2]{OCBE, Research Support Services, Oslo University Hospital, Norway}
\affil[3]{Department of Obstetrics, Division of Obstetrics and Gynecology, Oslo University Hospital Rikshospitalet, Oslo, Norway}
\affil[4]{Department of Medical Biochemistry, Oslo University Hospital}
\affil[5]{Institute of Clinical Medicine, Faculty of Medicine, University of Oslo }
\date{}

\begin{document}

\maketitle

\section*{Abstract}

High-dimensional interaction models are useful for studying, for example, how a large set of variables of interest, such as gene expression or other omics features, interact with a smaller set of modifying variables, such as clinical covariates. In this context, the pliable lasso has recently been proposed as an efficient method for screening large numbers of potential interaction terms under an asymmetric weak hierarchical constraint. In this work, we extend this framework by introducing PliableBVS, a Bayesian variable selection approach that preserves the hierarchical structure of the pliable lasso while inducing sparsity through spike-and-slab priors. The proposed model combines the continuous shrinkage effect of Bayesian lasso with a hierarchical spike-and-slab prior formulation that has two layers of decision variables: one governing the inclusion of main effects and another controlling the inclusion of interaction effects which is conditional on the inclusion of the corresponding main effects. This structure enables simultaneous selection of high-dimensional main and interaction effects within a coherent probabilistic framework. In simulation studies the proposed method outperforms the original pliable lasso in identifying active main and interaction effects, reducing false discoveries, and improving prediction accuracy in most scenarios. Applications with data from a labor onset study and a preeclampsia study demonstrate that PliableBVS selects biologically meaningful features and interactions.

\noindent\textbf{Keywords:} Bayesian variable selection, pliable lasso, spike-and-slab priors, interaction modeling, hierarchical sparsity.
\section{Introduction}
Recent advances in statistics and machine learning have substantially improved the potential for the use of high-dimensional data in medical research and healthcare applications, including in areas such as early disease detection \citep{skin}, patient risk stratification using electronic health records \citep{brom2020leveraging}, and biomarker discovery from high-dimensional omics data \citep{10.3389/fbinf.2024.1390607}. These developments enable more precise disease prediction, improved diagnosis, and personalized treatment strategies \citep{toh2021applications,habehh2021machine}.

In particular, the integration of omics data (genomics, proteomics, metabolomics, etc.) has revolutionized discovery of molecular biomarkers for disease prognosis, and diagnostic accuracy \citep{Zhao2021,Hu2011,ransohoff2003developing,doi:10.1200/JCO.2007.11.8497}. These advances improve our ability to predict health outcomes, tailor interventions, and improve patient care through data-driven decision-making. 

In many biomedical applications, however, the effect of such molecular biomarkers may vary across patient subgroups, due to for example differences in cancer type, sex, geographic location, or measurement time points (for longitudinal studies), necessitating the inclusion of interaction effects between high-dimensional omics variables and these subgroup-defining low-dimensional clinical or contextual covariates (the modifying variables).
%
\smallskip\\
\subsection*{Challenges of high‑dimensional interaction modeling}

Omics data are typically high-dimensional, where the number of measured variables can far exceed the number of samples. In such settings, identifying the relevant features requires effective variable selection techniques \citep{10.5555/2834535}. 
Penalized regression methods are widely used for this purpose,
with lasso \citep{tibshirani1996regression} being a prominent example that induces sparsity through the $L_1$-penalty. 

In addition to challenges such as  ensuring scalability in high-dimensional settings, and providing uncertainty quantification, modeling interactions between high-dimensional omics variables and a small set of mandatory subgroup-defining covariates as described above introduces challenges such as maintaining hierarchical constraints between main and interaction effects.
Several structured sparsity methods have been proposed to address these challenges by incorporating structured penalties that enforce hierarchical relationships while enabling the selection of main effects and interactions, for example the hierarchical lasso \citep{bien2013lasso} and the pliable lasso \citep{Tibshirani2020}. However, these approaches are typically formulated in a frequentist framework and do not directly provide measures of uncertainty for parameter estimates or for variable inclusion. To address these limitations, we propose a flexible structured sparsity Bayesian model with interactions, similar to the pliable lasso. 

Let $y\in \mathbb{R}^{N}$ denote the response vector, $X\in \mathbb{R}^{N\times p}$ the matrix of high-dimensional covariates to be subject to variable selection with entries $X_{ij}$, and $Z\in \mathbb{R}^{N\times K}$ the low-dimensional matrix of the mandatory modifying variables always included in the model  with entries $Z_{ik}$, $i=1,\ldots, N$, $k=1,\ldots,K$. Throughout this article, any vector is a column vector if there is no additional note. Let $X_j$ be the $j^{th}$ column of $X$ ($j=1,\ldots,p$), $Z_k$ the $k^{th}$ column of $Z$, and $\mathbf{1}$ an $N$-vector of ones. The general regression model with interaction effects between $X_j$ ($j=1,\ldots,p$) and $Z_k$ ($k=1,\ldots,K$) is 
\begin{equation}\label{m1}
y=\beta_{0}\mathbf{1}+Z\boldsymbol{\theta}_{0}+ X\boldsymbol{\beta}+\sum_{j=1}^p(X_{j}\circ Z)\boldsymbol{\theta}_{j}+\epsilon,
\end{equation}
where $\epsilon=(\epsilon_1,\cdots,\epsilon_N)^\top$ is the noise term assumed to be IID Gaussian, $\boldsymbol{\beta}\in \mathbb{R}^{p}$, $\theta\in\mathbb R^{p\times K}$ with $j^{th}$ row $\boldsymbol{\theta}_j$ and individual entries $\theta_{jk}$, $(X_{j}\circ Z)\in \mathbb{R}^{N\times K}$ is formed by multiplying each column of $Z$ componentwise
by the column vector $X_j$, $\beta_0$ is a scaler intercept, and $\boldsymbol{\theta}_0\in \mathbb{R}^{K}$. Note that we do not consider all second-order interactions among the covariates in $Z$, or among the covariates in $X$, as \cite{Agrawal2019}, because in many applications practitioners are mainly interested in the main effects of the (usually high-dimensional) omics matrix $X$ and their interactions with a small number of clinical or other patient phenotype variables that characterize the heterogeneity. 

The objective function for the linear pliable lasso model \citep{Tibshirani2020} following the above model is
\begin{equation}\label{p1}
\min_{\beta\in \mathbb{R}^{ p}, \theta\in \mathbb{R}^{p\times K}} \frac{1}{2N}  \left\Vert  y - \hat{y} \right\Vert _2^2
+   (1-\alpha)\lambda \sum_{j=1}^p (\| (\beta_j,\boldsymbol{\theta}_j)\|_2+\| \theta_j\|_2)  
+\alpha\lambda \sum_{j=1}^p\| \boldsymbol{\theta}_j\|_1, 
\end{equation}
where $\hat{y}=\hat{\beta}_{0}\mathbf{1}+Z\hat{\boldsymbol{\theta}}_{0}+ X\hat{\boldsymbol{\beta}}+\sum_{j=1}^p(X_{j}\circ Z)\hat{\boldsymbol{\theta}}_{j},
$ the $l_2$-norm penalty terms (appearing as overlapping groups) enforce group selection for each coordinate $j$ while maintaining the hierarchical property \citep{group,Simon01012013}, the $l_1$ penalty encourages sparsity within $\boldsymbol{\theta}_j$, and $\alpha\in [0,1]$ controls the relative weight of the two penalties. The hierarchical constraint is known as an asymmetric weak hierarchy, which ensures that $\theta_j$ can be nonzero only if $\beta_j$ is nonzero. More details about the weak asymmetric hierarchy with some biological examples can be found in \cite{Asenso15102025}. 
\smallskip\\
\subsection*{Motivation for a Bayesian formulation}

In the Bayesian framework, shrinkage is applied through priors for the regression coefficients \citep{Agrawal2019,bayeslaso}, either as continuous shrinkage priors or as spike-and-slab priors that introduce indicator variables to determine whether a feature should be included \citep{Carbonetto2012,George1993,BSGSS}. 
In this study, we introduce a spike-and-slab approach to incorporate both main and interaction effects in linear and logistic regression settings. Bayesian variable selection in hierarchical models has already been widely applied in biological contexts, for example to group genes within the same biological pathway \citep{10.1214/19-AOAS1271}, or to model interactions between biological variables \citep{liu2015bayesian,vandeWielAmestoyHoogland+2024,Zhao2021, stingo2011bayesian}. These Bayesian approaches do not only improve the interpretability of the models through the selection of a small panel of features, but also enable the production of uncertainity quantification for estimates via posterior distributions—something that standard non-Bayesian penalized regressions do not provide. 

In this paper, we propose a Bayesian approach to modeling main and interaction effects that mirrors the pliable lasso \citep{Tibshirani2020}. 
To enforce the asymmetric weak hierarchy, we introduce two latent decision parameters, each modeled as a probability drawn from a Beta distribution; one governs the inclusion of a main effect and the other governs the inclusion of its corresponding interaction terms.
These parameters act as stochastic gates that decide whether each effect should be retained in the regression. Specifically, the inclusion probability of an interaction can depend on that of its main effect, thereby enforcing the desired hierarchy.
This formulation provides a flexible way to control shrinkage and variable selection at multiple levels of the hierarchy, a concept that has been widely explored in hierarchical shrinkage priors \citep{10.1214/15-BA990,BSGSS,Chen02072016,vandeWielAmestoyHoogland+2024}. Each feature and its interactions are given zero-inflated mixture priors in the normal-Dirac form of the spike-and-slap prior \citep{Mitchell01121988,Chen02072016,BSGSS}.  
We compare our Bayesian method with the non-Bayesian pliable lasso, which handles the same modeling problem. An open-source R software package \textbf{PliableBVS} \citep{PliableBVSpaper} implementing the proposed methodology is available, and we also provide a language-agnostic API \href{https://bpliable-api.onrender.com/__docs__/}{PliableBVS-api} to allow non-R users to run the analysis and generate plots.   


The remainder of the paper is organized as follows. Section \ref{method} introduces the proposed Bayesian pliable lasso model and describes the MCMC algorithm, including an extension to binary outcomes. Section \ref{simulation} presents simulation studies comparing the proposed method with the pliable lasso for both continuous and binary outcomes. Section \ref{real} illustrates the proposed approach using two real data applications involving labor onset and preeclampsia. The primary objective of these analyses is feature discovery and prediction, and any biological interpretations of the identified variables and interactions should be regarded as hypothesis-generating rather than causal. Section \ref{conclusion} concludes the paper.

\section{Methods}\label{method}

\subsection{Prior setup}
\subsubsection{Linear regression}
When only a few of the $p$ high-dimensional predictors are truly associated with the response, a zero-inflated mixture (spike-and-slab) prior is placed on each main-effect coefficient $\beta_{j}$, $j=1,\cdots,p$ to induce variable selection \citep{George1993,BSGSS}. We propose the Bayesian pliable lasso for the linear regression case;

\begin{gather}
y | \beta_0, \boldsymbol{\beta}, \boldsymbol{\theta}_0, \boldsymbol{\theta}, \sigma^2 \sim \mathcal{N}(\beta_0\mathbf{1}+Z\boldsymbol{\theta}_0+X\boldsymbol{\beta}+\sum_{j=1}^p(X_j\circ Z)\boldsymbol{\theta}_j,\sigma^2\mathbb{I}_N)\label{y_cont},\\
\beta_0 \sim \mathcal{N}(0,c^2),\nonumber\\
\boldsymbol{\theta}_0 \sim \mathcal{N}(\bm 0,v^2\mathbb{I}_K), \nonumber\\
\sigma^2\sim \text{Inverse Gamma}(\lambda_1,\lambda_2), 
\nonumber\\
\beta_{j}|\tau_{\beta_j}^2,\sigma^2 \sim (1-\rho)\mathcal{N}(0,\ \tau_{\beta_j}^2\sigma^2) + \rho\delta_0(\beta_{j}), \nonumber \\
\tau_{\beta_j}^2 \sim  \text{Gamma}(1, (1-\alpha)^2\lambda^2/2) \nonumber
\label{formula:spike_slab}
\end{gather}
where $\text{Gamma}(1, x)=\exp(x)$ which is a Laplace prior, equivalent to the Bayesian lasso variance, $c$ and $v$ are  non-zero constants, $\alpha \in [0,1]$ and plays same role as the $\alpha$ in pliable lasso, and $\delta_0(\cdot)$ is the Dirac delta function at 0. 
Similar to the pliable lasso model in (\ref{p1}) with asymmetric weak hierarchy, we again allow selection of the interaction effect $\theta_{jk}$ only if the corresponding main effect $\beta_j$ is selected, by assuming a multivariate zero-inflated mixture prior for $\theta_{jk}$ and using independent spike-and-slab priors for the coefficients of interaction terms as
\begin{gather} 
\theta_{jk}| \beta_{j}\neq 0, \tau_{\theta_{jk}}^2,\sigma^2 \sim (1-\zeta) \mathcal{N}(0,\ \tau_{\theta_{jk}}^2\sigma^2) + \zeta\delta_0(\theta_{jk}), \nonumber  \\
\tau_{\theta_{jk}}^2 \sim  \text{Gamma}(1, \alpha^2\lambda^2/2).
\tag{4}\label{formula:spike_slab2}
\end{gather}
Conditioning on $\beta_{j}\neq 0$
 enforces an asymmetric weak hierarchy, ensuring that interaction effects are included only when the corresponding main effect is active. For the inclusion probabilities $\rho$ and $\zeta$, we assign beta priors:
\begin{align*}
\rho &\sim \text{Beta}(a_\rho,b_\rho), \\
\zeta &\sim \text{Beta}(a_\zeta,b_\zeta),   
\end{align*}
where $a_\rho,b_\rho,a_\zeta,b_\zeta>0$.


\subsubsection{Logistic regression}
The same hierarchical prior can be used when the response $y$ is binary.
Following the P\'olya-gamma augmentation of \cite{Polson_2013}, we introduce a latent Gaussian working response $O=(O_1,\ldots,O_N)$ and write
\begin{gather}
O| \beta_0, \boldsymbol{\beta}, \boldsymbol{\theta}_0, \boldsymbol{\theta}, \Omega \sim \mathcal{N}(\beta_0\mathbf{1}+Z\boldsymbol{\theta}_0+X\boldsymbol{\beta}+\sum_{j=1}^p(X_j\circ Z)\boldsymbol{\theta}_j,\Omega^{-1} )\label{o_bin},\\
\omega_i \sim \text{PG}(1,0), 
\label{formula:spike_slab2}
\end{gather}
where $ \Omega=diag(\omega_i\ldots, \omega_N)$, \quad $O=(O_1,\ldots, O_N), \quad O_i=(y_i-1/2)/\omega_i, \quad i=1, \ldots, N$.
In the formulation above, $\text{PG}(a,b)$ is P\'olya-gamma distribution. All other prior setup remains the same as in the linear regression case.  Details of the P\'olya-gamma distribution and how it can be used to implement the logistic regression are given by \cite{Polson_2013}. In short, they showed that the likelihood of the logistic regression can be represented as a mixture of Gaussians based on a P\'olya-gamma distribution by applying the following principle: 
\begin{equation}
    \begin{split}
        \mathcal{L}_i(\beta)&=\frac{(e^{X_i^T\beta} )^{y_i}}{(1+e^{X_i^T\beta} )} \propto e^{\kappa_iX_i^T\beta} \int_{0}^{\infty} e^{-\omega_i (X_i^T\beta)^2/2} \pi(\omega_i|1,0) d\omega_i,
    \end{split}\label{poly_theorem}
\end{equation}
where $\kappa_i=y_i-1/2$ and $\pi(\omega_i|1,0)$ represents the density of a P\'olya-gamma random variable $\omega_i$ with parameters $(1,0)$. 
Conditional on $\omega_i$, the likelihood (\ref{poly_theorem}) becomes proportional to a Gaussian distribution, i.e., 
$\mathcal{L}_i(\beta|\omega_i) \propto \exp\{\kappa_iX_i^T\beta -\omega_i (X_i^T\beta)^2/2\}$.

\subsection{Gibbs sampler}

\subsubsection{Linear regression case}
Given the hierarchical model described above, we derive a Gibbs sampler for posterior inference. The Gibbs sampler proceeds by iteratively sampling from the following full conditional distributions;

\begin{multline}\label{gibbs}
p(\beta_0,\boldsymbol{\theta}_0,\boldsymbol{\beta},\boldsymbol{\theta},\sigma^2,
\bm\tau_{\beta}^2,\bm\tau_{\theta}^2, \rho,\zeta | X,Z,y)\\
\propto \text{Likelihood}(y |X, Z, \beta_0,\boldsymbol{\theta}_0,\boldsymbol{\beta},\boldsymbol{\theta},\sigma^2,
\bm\tau_{\beta}^2,\bm\tau_{\theta}^2,\rho,\zeta )\times\\ \pi(\beta_0)\pi(\boldsymbol{\theta}_0|...) \pi(\boldsymbol{\beta}|...)\pi(\boldsymbol{\theta}|...)  
 \pi(\bm\tau_{\beta}^2)\pi(\bm\tau_{\theta}^2) \pi(\bm\rho)\pi(\bm\zeta) \pi(\sigma^2)\\
= (\sigma^2)^{-\frac{N}{2}} \text{exp} \Biggl \{ -\frac{1}{2\sigma^2} (y-[\beta_0\mathbf{1}+Z\boldsymbol{\theta}_0+X\boldsymbol{\beta}+\sum_{j=1}^p(X_j \circ Z)\boldsymbol{\theta}_j])^T(y-[\beta_0\mathbf{1}+Z\boldsymbol{\theta}_0+X\boldsymbol{\beta}+\sum_{j=1}^p(X_j \circ Z)\boldsymbol{\theta}_j]) \Biggr\}\\
\
\times (2\pi c^2)^{-\frac{1}{2}} \text{exp}\biggl\{-\frac{1}{2c^2}(\beta_0)^2\biggr\} \ \times \ (2\pi v^2)^{-\frac{K}{2}} \text{exp}\biggl\{-\frac{1}{2v^2}\boldsymbol{\theta}_0^T\boldsymbol{\theta}_0\biggr\}\\
\
\times \prod_{j=1}^p \Biggl[ {(1-\rho)} \biggl\{\frac{1}{ \sqrt{ 2\pi\sigma^2\tau_{\beta_j}^2  }} \text{exp} \biggl(-\frac{\beta_j^2}{2\sigma^2\tau_{\beta_j}^2} \biggr) \biggr\}I[\beta_j\neq 0] \ + \ \rho\delta_0(\beta_j) \Biggr]\\
\
\times \prod_{j=1}^p \prod_{k=1}^K \Biggl[ I(\beta_j\neq 0) \left( (1-\zeta) \biggl\{\frac{1}{ \sqrt{ 2\pi\sigma^2\tau_{\theta_{jk}}^2  }} \text{exp} \biggl(-\frac{\theta_{jk}^2}{2\sigma^2\tau_{\theta_{jk}}^2} \biggr) \biggr\}I[\theta_{jk}\neq 0]  \ + \ \zeta\delta_0(\theta_{jk})      \right) \Biggr]\\
\
\
\times \prod_{j=1}^p ((1-\alpha)^2\lambda^2)(\tau_{\beta_j}^2)^{(1-1)} \text{exp}\biggl(-\frac{(1-\alpha)^2\lambda^2}{2}\tau_{\beta_j}^2 \biggr)\\
\
\times \prod_{j=1}^p  \prod_{k=1}^K (\alpha^2\lambda^2)(\tau_{\theta_{jk}}^2)^{(1-1)} \text{exp}\biggl(-\frac{\alpha^2\lambda^2}{2}\tau_{\theta_{jk}}^2 \biggr)
\
\ 
\\
\
\times  \rho^{a_\rho-1}(1-\rho)^{b_\rho-1}
\ 
\times \zeta^{a_\zeta-1}(1-\zeta)^{b_\zeta-1}
\ 
\times (\sigma^2)^{-\lambda_1-1} \text{exp}\biggl(-\frac{\lambda_2}{\sigma^2}\biggr),
\end{multline}
where $\bm\tau_{\beta}=(\tau_{\beta_1},\ldots, \tau_{\beta_p})$, $\bm\tau_{\theta}=(\tau_{\theta_{1}},\ldots, \tau_{\theta_{p}})$, with each $\tau_{\theta_{j}}=(\tau_{\theta_{k1}},\ldots,\tau_{\theta_{jK} }  )$. 
Below we give the full conditional distributions that are sampled in each Gibbs iteration.

\begin{itemize}
\item  {\bf Step 1 (Gibbs for main effects $\boldsymbol{\beta}$ and variable selection):} \\
Let $\beta_{-j}$ be the coefficients of the main effects without the $j$-th component, i.e. $\beta_{-j}=(\beta_1,\beta_2,\cdots,\beta_{j-1},\beta_{j+1},\cdots, \beta_p)$ and $X_{-j}$  be the covariate matrix without the $\text{j}^{\text{th}}$ column. Also, let $\bar{y}=y-\beta_0\mathbf{1}-Z\boldsymbol{\theta}_0-\sum_l(X_l \circ Z)\boldsymbol{\theta}_l$ and $\mu_j=S_jX_j^T(\bar{y}-X_{-j}\beta_{-j})$, $S_j=(X_j^TX_j+\frac{1}{\tau_{\beta_j}^2})^{-1}$. Then the conditional posterior distribution of $\beta_j$ is a spike-and-slab distribution 

\begin{equation}\label{batejupdate}
    \beta_j|\ldots \sim (1-\eta_j)\mathcal{N}(\mu_j,\sigma^2S_j)+\eta_j\delta_0(\beta_j), \ \ j=1,\cdots,p, 
\end{equation}
where
\begin{equation}\label{l_j_update}
    \eta_j=p(\beta_j=0|\ldots)=\frac{\rho}{\rho+(1-\rho)(\tau_{\beta_j}^2)^{-\frac{1}{2} } S_j^{\frac{1}{2}} \text{exp}\biggl\{ \frac{1}{2\sigma^2}  ( S_j (X_j^T (\bar{y}-X_{-j}\beta_{-j})  )^2 ) \biggr\} },
\end{equation}

\begin{equation}\label{rhoupdate}
    \rho|\ldots \sim \text{Beta} \Biggl(a_{\rho} + p-\sum_j Q_j, b_\rho + \sum_jQ_j  \Biggr)
\end{equation}
where 
\begin{equation}\label{Qj}
    Q_{j}=\begin{cases}
    1, & \text{if} \ \beta_{j}\neq 0,\\
    0, & \text{if} \ \beta_{j} = 0.
  \end{cases}
\end{equation}

\item {\bf Step 2 (Gibbs for $\tau_{\beta_j}^2$ update):}\\
Let $\iota_j^2=\frac{1}{\tau_{\beta_j}^2},\ \ j=1,\cdots,p$. Then 
\begin{equation}\label{taujupdate}
    \iota_j^2|\ldots \sim \begin{cases}
    \text{Inverse Gamma} \bigg(\text{shape}=1,\text{scale}=\frac{(1-\alpha)^2\lambda^2}{2}\bigg), & \text{if} \ \beta_j=0,\\
    \text{Inverse Gaussian} \bigg(\text{mean}=\frac{(1-\alpha)\lambda\sigma}{|\beta_j|},\text{shape}=(1-\alpha)^2\lambda^2 \bigg), & \text{if} \ \beta_j\neq0.
  \end{cases}
\end{equation}
This update arises from the normal–gamma mixture representation of the lasso prior.


\item {\bf Step 3 (Gibbs for the interaction effects $\boldsymbol{\theta}$ and variable selection):}\\ For the $\theta_{jk}$ update we define $\bar{y}$ as $\bar{y}=y-\beta_0\mathbf{1}-Z\boldsymbol{\theta}_0-X\boldsymbol{\beta}$ and let $\mu_{jk}=S_{jk}[X_j \circ Z_k]^T(\bar{y}-\sum_{w\neq k}X_j \circ Z_w\theta_{jw})$, $S_{jk}=([X_j \circ Z_k]^T[X_j \circ Z_k]+\frac{1}{\tau_{\theta_{jk}}^2})^{-1}$. Then the conditional posterior distribution of $\theta_{jk}$ given that $\beta_j\neq 0$ is a spike-and-slab distribution 

\begin{equation}\label{theta_j_upate}
    \theta_{jk}| \beta_{j}\neq 0, \ldots \sim (1-\gamma_{jk})\mathcal{N}(\mu_{jk},\sigma^2S_{jk})+\gamma_{jk}\delta_0(\theta_{jk}),
\end{equation}
where
\begin{equation}\label{r_jk_update}
    \gamma_{jk}=p(\theta_{jk}=0|\ldots)=\frac{\zeta}{\zeta+(1-\zeta)(\tau_{\theta_{jk}}^2)^{-\frac{1}{2} } S_{jk}^{\frac{1}{2}} \text{exp}\biggl\{ \frac{1}{2\sigma^2}  \mu_{jk}^2 / S_{jk}  \biggr\} },
\end{equation}

\begin{equation}\label{zetapdate}
    \zeta|\ldots \sim \text{Beta} \Biggl(a_{\zeta} + \sum_{j:\beta_j \neq 0} K-\sum_{j:\beta_j \neq 0} \sum_{k} R_{jk}, b_\zeta + \sum_{j:\beta_j \neq 0}\sum_k R_{jk}  \Biggr),
\end{equation}
 where 
\begin{equation}\label{Qj}
    R_{jk}=\begin{cases}
    1, & \text{if} \ \theta_{jk}\neq 0,\\
    0, & \text{if} \ \theta_{jk} = 0.
  \end{cases}
\end{equation}

\item {\bf Step 4 (Gibbs for $\tau_{\theta_{jk}}^2$ update):}\\
Let $\nu_{jk}^2=\frac{1}{\tau_{\theta_{jk}}^2}, \ j=1,\cdots,p, \ k=1,\cdots,K$. Then 

\begin{equation}\label{taujkupdate}
    \nu_{jk}^2|\ldots \sim \begin{cases}
    \text{Inverse Gamma} \bigg(\text{shape}=1,\text{scale}=\frac{\alpha^2\lambda^2}{2}\bigg), & \text{if} \ \theta_{jk}=0,\\
    \text{Inverse Gaussian} \bigg(\text{mean}=\frac{\alpha\lambda\sigma}{|\theta_{jk}|},\text{shape}=\alpha^2\lambda^2 \bigg), & \text{if} \ \theta_{jk}\neq0.
  \end{cases}
\end{equation}

\item {\bf Step 5 (Gibbs for $\sigma^2$ ):}\\ The  conditional posterior distribution of  $\sigma^2$ is conditionally inverse gamma, i.e., 

\begin{multline}\label{sigma}
    \sigma^2 | \ldots \sim \text{Inverse Gamma} \biggr( \frac{N}{2} + \frac{\sum_jQ_j}{2} +\frac{\sum_{jk}R_{jk}}{2} +\lambda_1, \\ 
    \frac{1}{2}\biggl[ (y-\beta_0\mathbf{1}-Z\boldsymbol{\theta}_0-X\boldsymbol{\beta}-\sum_j (X_j \circ Z)\boldsymbol{\theta}_j)^T(y-\beta_0\mathbf{1}-Z\boldsymbol{\theta}_0-X\boldsymbol{\beta}-\sum_j (X_j \circ Z)\boldsymbol{\theta_j})\\+ \boldsymbol{\beta}^T D_{\tau_\beta}^{-1} \boldsymbol{\beta} +\sum_j\boldsymbol{\theta}_j^T D_{\tau_{\theta_j} }^{-1}\boldsymbol{\theta}_j    +2\lambda_2\biggr]  \biggl),
\end{multline}
\textcolor{red}{
}
where $D_{\tau_\beta}=\text{diag}\{\tau_{\beta_1},\cdots, \tau_{\beta_p} \}$ and $D_{\tau_1} \in \mathcal{R}^{K \times K \times p}$ such that $D_{\tau_{\theta_j}}=\text{diag}\{\tau_{\theta_{j1}},\cdots, \tau_{\theta_{jK}}\}, \ j=1,\cdots, p $.


\end{itemize}
The global penalty parameter $\lambda$ is estimated using an empirical Bayes approach. By this, we estimate $\lambda$ from the data through marginal maximum likelihood using a Monte Carlo EM algorithm (\cite{Casella2001EmpiricalBG}; \cite{bayeslaso}; \cite{BSGSS}) to estimate $\lambda$. The $s^{\text{th}}$ update is 

\begin{equation}
    \sqrt{\frac{2p+2(p\times K)}{(1-\alpha)^2\sum_j\mathbf{E}_{\lambda^{(s-1)}} [\tau_{\beta_j}^2|y] + \alpha^2 \sum_j\sum_k \mathbf{E}_{\lambda^{(s-1)}} [\tau_{\theta_{jk}}^2| y]}},
\end{equation}
where we will replace the conditional expectations $\tau_{\beta_j}^2$ and $\tau_{\theta_{jk}}^2$ with their sample averages from the
Gibbs sampler run based on $\lambda^{(s-1)}$.

Note that we use MCMC to estimate $\boldsymbol{\beta}$ and $\boldsymbol{\theta}$ sequentially, which is more scalable than some optimization methods such as ADMM that estimates $\boldsymbol{\beta}$ and $\boldsymbol{\theta}$ simultaneously and operates them as one matrix. 

\subsubsection{Logistic regression}
Applying the P\'olya-gamma augmentation \eqref{poly_theorem} and the same hierarchical priors as in the linear case, the joint posterior distribution of the logistic regression is 
\begin{multline}\label{gibbs2}
p(\beta_0,\boldsymbol{\theta}_0,\boldsymbol{\beta},\boldsymbol{\theta},\sigma^2,\Omega,
\bm\tau_{\beta}^2,\bm\tau_{\theta}^2, \rho,\zeta | X,Z,O)\\
\propto \text{Likelihood}(O |X, Z, \beta_0,\boldsymbol{\theta}_0,\boldsymbol{\beta},\boldsymbol{\theta},\sigma^2,\Omega,
\bm\tau_{\beta}^2,\bm\tau_{\theta}^2,\rho,\zeta )\times\\ \pi(\beta_0)\pi(\boldsymbol{\theta}_0|...) \pi(\boldsymbol{\beta}|...)\pi(\boldsymbol{\theta}|...) 
 \pi(\bm\tau_{\beta}^2)\pi(\bm\tau_{\theta}^2) \pi(\bm\rho)\pi(\bm\zeta)\pi(\Omega) \pi(\sigma^2)\\
=  \text{exp} \Biggl \{ -\frac{1}{2} (O-[\beta_0\mathbf{1}+Z\boldsymbol{\theta}_0+X\boldsymbol{\beta}+\sum_{j=1}^p(X_j \circ Z)\boldsymbol{\theta}_j])^T\Omega(O-[\beta_0\mathbf{1}+Z\boldsymbol{\theta}_0+X\boldsymbol{\beta}+\sum_{j=1}^p(X_j \circ Z)\boldsymbol{\theta}_j]) \Biggr\}\\
\
\times \prod_{i=1}^N \pi(\omega_i) \\
\times \ldots
\ 
\times (\sigma^2)^{-\lambda_1-1} \text{exp}\biggl(-\frac{\lambda_2}{\sigma^2}\biggr).
\end{multline}
In the logistic regression setting, $\sigma^2$ acts as a global shrinkage parameter rather than a noise variance. We  show the conditional posterior for $\beta_0,\theta_0,\beta,\theta,\Omega^2$ and $\sigma^2$ in the suplemetary section \ref{suplement}. The conditional posteriors for the other parameters are exactly the same as in the linear case. 

\section{Simulations}\label{simulation}

We conducted simulation studies to evaluate the performance of the proposed Bayesian pliable lasso (PliableBVS) in comparison with the frequentist pliable lasso (PliableLasso). Performance was assessed in terms of variable selection accuracy and predictive performance using independently generated test datasets.

\subsection{Simulation under linear regression}

We consider five scenarios for simulations based on linear regression.
Scenarios 1-2 are in low-dimensional settings with low signal-to-noise ratio (SNR), one with a continuous modifying variable $Z$ and the other with binary $Z$. 
Scenarios 3-4 are similar to scenarios 1-2 but with high SNR, in which scenario 4 is high-dimensional. 
Scenario 5 is low-dimensional with moderate SNR. The predictors 
$X$ were generated independently from a standard normal distribution unless otherwise stated. More details are as follows.

\paragraph{Scenarios 1-2.} We generated data with $N=200$, $p=20$, $K=4$ and standard Gaussian independent predictors. The data-generating mechanism follows that of \cite{Tibshirani2020}:
\begin{equation}\label{sim_model1}
y=X_1\beta_{1}+X_2\beta_{2}+X_3(\beta_{3}\mathbf{1}+2Z_1)+X_4\beta_{4}(\mathbf{1}-2Z_2)+\varepsilon,
\end{equation}
where $\boldsymbol{\beta}=(2, -2, 2, 2,0,\ldots)$. In the first scenario,  $Z$ is a matrix $N\times K$  drawn from the standard normal distribution with zero mean and 1 as standard deviation and $\varepsilon \sim \mathbb{N}(0,\sigma I_N)$. The signal-to-noise ratio (SNR) was about 2. In the second scenario, $Z$ is drawn from the Bernoulli distribution with probability of 0.5. 

\paragraph{Scenarios 3-5.} The data-generating mechanism is 
\begin{equation}\label{sim_model2}
y=\sum_{j=1}^p X_{j}\beta_{j}\mathbf{1}+\sum_{j=1}^p(X_{j}\circ Z)\boldsymbol{\theta}_{j}+\epsilon,
\end{equation}
where the coefficients of the interaction part in the above model are set up as 
\begin{equation}\label{G_matrix}
        \begin{bmatrix}
        & \theta_{j1} & \theta_{j2} & \theta_{j3} & \theta_{j4} \\
	\theta_1 & 0.5 & 0 & 0 & -0.5\\
	 \theta_2 & 0 & 0 & -0.5& 0\\
	\theta_3 & 0 & 0 & 0 & 0\\
       \theta_4 &   0 & 0.5 & 0 & 0\\
	\theta_5 & 0 & 0 & 0& 0\\
        \theta_{6} & 0 & -0.5 & 0 & 0.5\\
         \theta_{7} & 0 & 0 & 0 & 0\\
        \theta_{8} &0.5 & 0 & 0.5 & 0\\
        & \vdots & \vdots & \vdots & \vdots\\
	\end{bmatrix}
  \end{equation}
We have three simulation scenarios (scenarios 3, 4 and 5) under this model. The regression coefficient vector is defined as $\boldsymbol{\beta}=(0.9,-0.7,0.6,0.8,0.4,\mathbf{0.8},0,\ldots,0)$, with $\mathbf{0.8}$ a vector of length 5 and the total length of $\boldsymbol{\beta}$ is $p=20$ (scenario 3 and 5) or $p=500$ (scenario 4). For scenarios 3 and 4 the signal-to-noise ratio was chosen to be about 6, while for scenario 5 it was chosen to be about 2.

\begin{figure}[hbt!]

  \includegraphics[height=7.5cm,width=.8\textwidth]{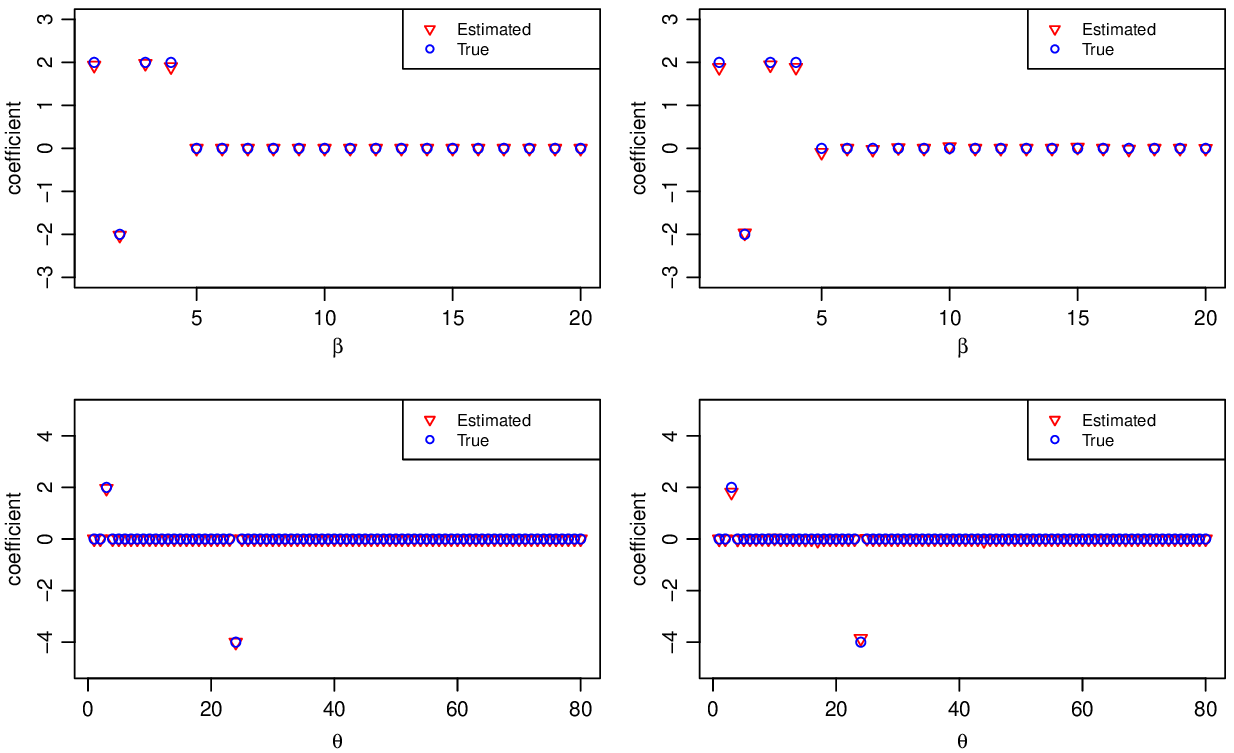}
 
\caption{Plots for main (top row) and interaction (bottom row) effects for linear regression simulation scenario 1 for PliableBVS (left column) and PliableLasso (right column). }\label{plot1}
\end{figure}

\begin{figure}[ht]
\begin{subfigure}{.5\textwidth}
  \centering
  \includegraphics[height=5cm,width=\textwidth]{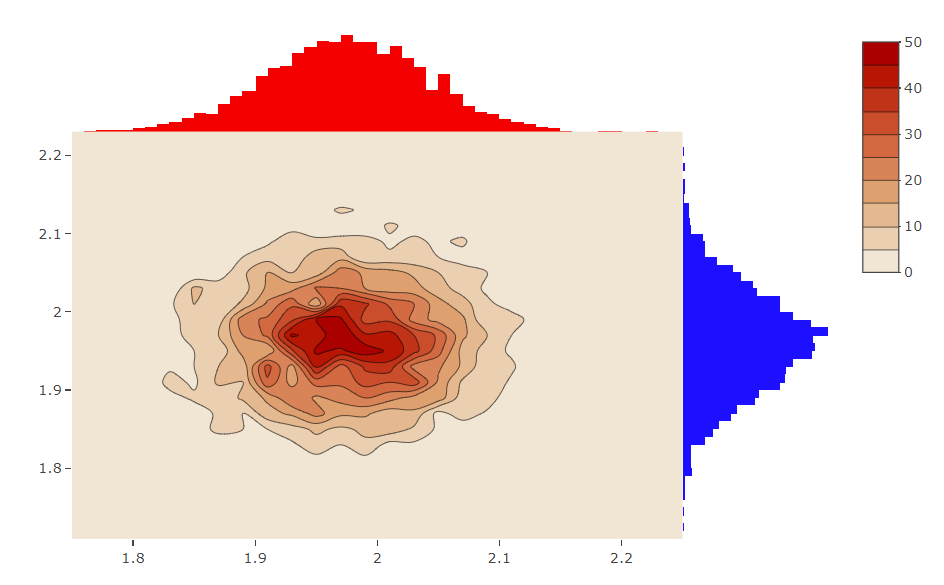}
 \caption{}
\end{subfigure}
\begin{subfigure}{.5\textwidth}
  \centering
  \includegraphics[height=5cm,width=\textwidth]{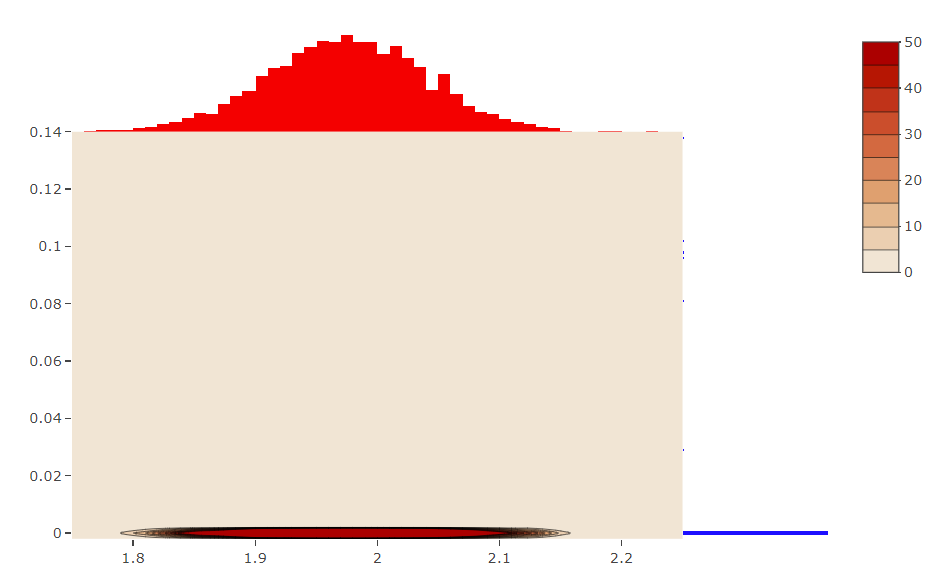}
 \caption{}
\end{subfigure}
\begin{subfigure}{.5\textwidth}
  \centering
  \includegraphics[height=5cm,width=\textwidth]{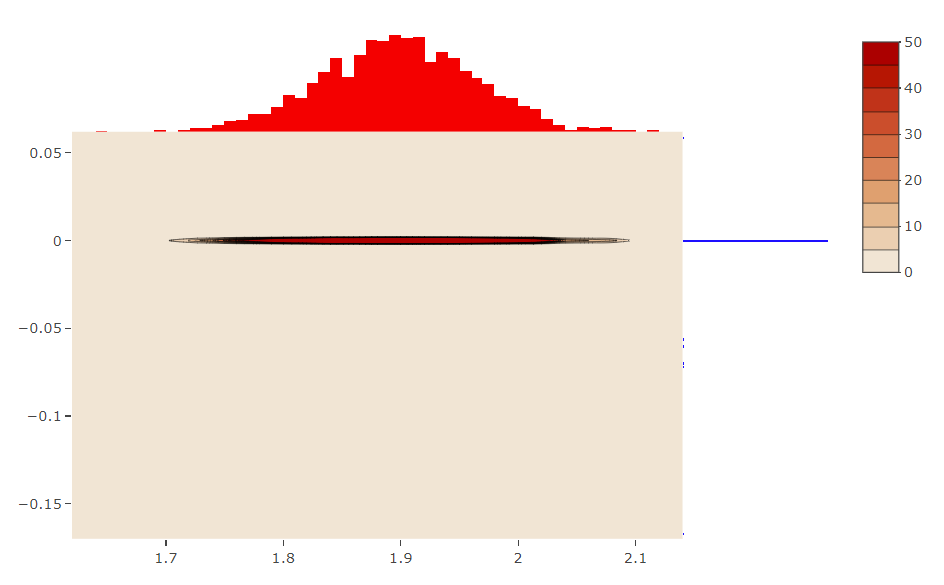}
 \caption{}
\end{subfigure}
\begin{subfigure}{.5\textwidth}
  \centering
  \includegraphics[height=5cm,width=\textwidth]{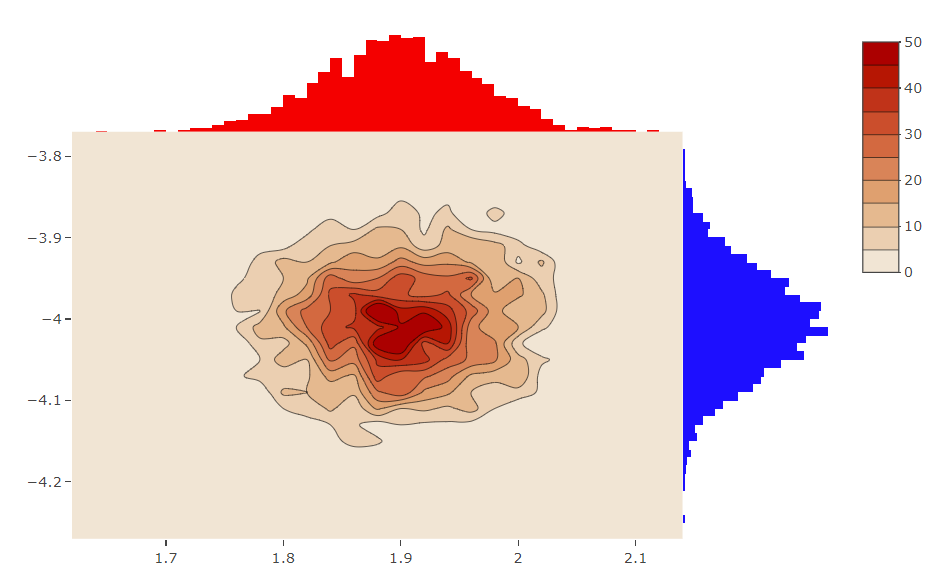}
 \caption{}
\end{subfigure}
\caption{Contour plots for the main and  interaction effects for linear regression simulation scenario 1;  (a) $j=3, k=1$ , (b)  $j=3, k=2$ (c) $j=4, k=1$ (d) $j=4, k=2$. The red histograms represent the marginal posterior of main effects $\beta_j$; the blue represent the marginal posterior of the corresponding interaction effects $\theta_{jk}$.  }\label{plot4}
\end{figure}

\begin{figure}[ht]

  \centering
  \begin{subfigure}{.5\textwidth}
  \includegraphics[height=5cm,width=\textwidth]{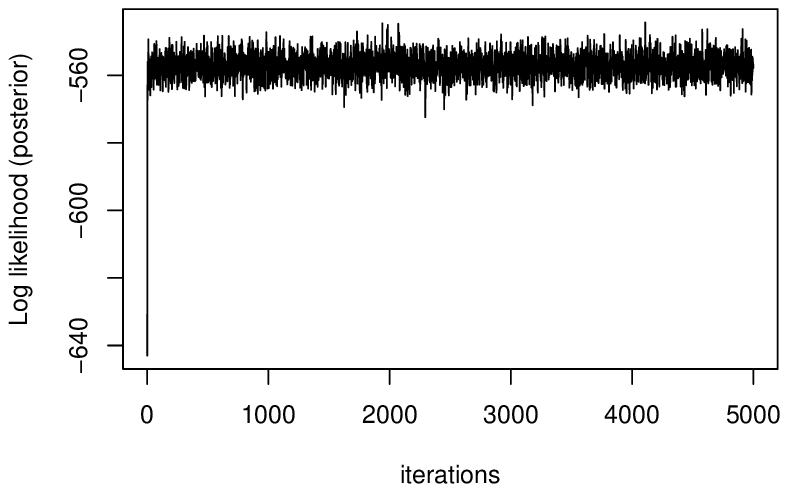}
 \caption{}
\end{subfigure}%
   \begin{subfigure}{.5\textwidth}
  \includegraphics[height=5cm,width=\textwidth]{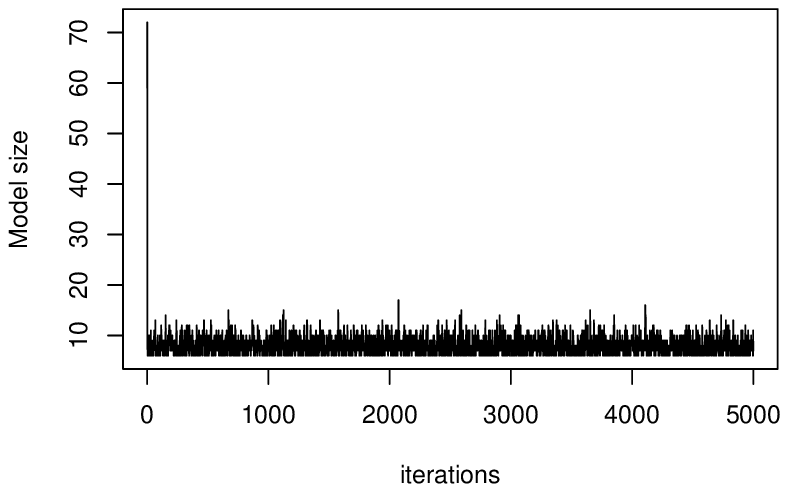}
 \caption{}
\end{subfigure}%

\caption{Diagnostic plots of the MCMC sampler for linear regression simulation scenario 1. }\label{trace}
\end{figure}

For each of the five simulation scenarios, we run the experiment 50 times, with different training data and different random seeds,  and compare our method with the pliable lasso. All works were carried out using R version 4.1.3. We set $\alpha=0.5$ for both our method (``PliableBVS'') and the pliable lasso (``PliableLasso''). For all simulations and real data analysis, we used the pliable R package (version 1.1.1; available from the CRAN archive) with its default parameter settings except those that were stated otherwise. The tuning parameter $\lambda$ for PliableLasso was selected using cross-validation, while that of PliableBVS was selected using the empirical Bayes approach described earlier. We use 5000 iterations with 2000 burn-in for our PliableBVS.  Our MCMC iterations indicated to reaching its stationary distribution as seen in the trace plots (figure \ref{trace}). The summary of all simulations is shown in Table \eqref{table2}. Overall, the results of the simulation experiments can be summarized as follows:
\begin{itemize}
    
    \item In all scenarios and across the different SNR settings, PliableBVS consistently achieves lower test MSE and improved variable selection accuracy across all scenarios, indicating more effective recovery of the true underlying model structure.
    \item The Bayesian marginal posterior distributions for the elements of $\beta$ and $\theta$ all appear to be unimodal (see figure \ref{plot4} for examples), indicating stable estimation. The joint posterior distributions of a main effect and its associated interaction do not indicate strong correlation. 

    \item Both PliableBVS and the PliableLasso are able to identify the true non-zero coefficients ( figure \ref{plot1}). However, PliableBVS consistently achieved lower false positive rates than PliableLasso across the simulation settings considered here. This may be related to the tendency of cross-validation-based tuning procedures for lasso-type models to favor larger models and over-select variables in prediction-oriented settings \citep{Shao1993,Homrighausen2013}. The introduction of the spike-and-slab prior in PliableBVS induces stronger shrinkage for true zero coefficients while at the same time allowing less shrinkage for true non-zero coefficients, as opposed to the frequentist pliable lasso, which enforces the same shrinkage on all coefficients, in line with well-established model consistency properties of the lasso-type penalty employed \citep{leng_lasso,zhao2006model}. In addition, PliableBVS uses the median probability model (MPM) \citep{10.1214/009053604000000238} to determine a single model with selected variables, which includes all variables whose marginal inclusion probabilities exceed 0.5. This approach provides a parsimonious yet stable model that balances selection uncertainty and predictive performance.
    \item PliableBVS also outperforms PliableLasso  in estimating the true values of both main effects and interaction effects, represented as $(1/(p+p\times K))\lVert \hat{\mathbf{B}}-\mathbf{B}\rVert_1$ in table \ref{table2}, where $\mathbf{B}$ is defined as $[\beta,\theta]$.
    \item In scenarios 3 and 4 with a high SNR, PliableBVS achieved higher sensitivity than PliableLasso when the number of predictors $p$ was large, and comparable sensitivity when $p$ was small.
In contrast, in the same simulation example under low SNR (scenario 5), PliableBVS showed slightly lower sensitivity than PliableLasso.
Under the same settings, PliableLasso exhibited poor specificity for both small and large $p$, thus resulting in the larger proportion of false positives mentioned above.
\end{itemize}

\begin{table}[ht]
\small \small \small
\centering
  \begin{center}
    \caption{Results from the linear regression simulation examples. }
    \label{table2}
    \begin{threeparttable}
    \resizebox{\textwidth}{!}{%
    \begin{tabular}{|l l l l l l l l|} 
    \hline
     \textbf{Model} & $(1/(p+p\times K))\lVert \hat{\mathbf{B}}-\mathbf{B}\rVert_1$\tnote{1*} & Sensitivity\tnote{2*} & Specificity\tnote{3*} & Accuracy\tnote{4*} & FPR\tnote{*} &\# Non-zeros\tnote{5} & Test error \tnote{6*} \\
      \hline
      Scenario 1 &$p=20$, $SNR=2$  & & & & & & \\
     PliableBVS & 0.021 (0.010) & 0.993 (0.047) & 0.997 (0.007) & 0.996 (0.007) &  0.003 (0.007)& 4 (2) & 1.663 (2.824) \\
     PliableLasso  & 0.059 (0.023)& 1.000 (0.000) & 0.743 (0.133) & 0.758 (0.125) & 0.257 (0.133)&12 (17) & 3.127 (1.142) \\
  
Scenario 2 & $p=20$ , $SNR=2$ & & & & & &\\
     PliableBVS  &0.021 (0.011) & 0.980 (0.054) & 0.998 (0.005) & 0.997 (0.005) & 0.002 (0.005) & 4 (2) & 0.773 (0.492)\\
     PliableLasso & 0.097 (0.026) &1.000 (0.000) & 0.659 (0.137) & 0.680 (0.128) & 0.341 (0.137) & 16 (22) & 1.839 (0.611)  \\
     Scenario 3  &$p=20$, $SNR=6$ & & & & & &\\ 
     PliableBVS  & 0.018 (0.005) & 1.000 (0.008) &0.970 (0.024) &  0.975 (0.021) & 0.030 (0.024) & 11 (8) &0.265 (0.119)  \\ 
     PliableLasso   & 0.034 (0.007) & 1.000 (0.000) & 0.530 (0.140) & 0.615 (0.115) &  0.469 (0.140) & 17 (39) & 0.459 (0.147)   \\ 
     Scenario 4 & $p=500$, $SNR=6$ & & & & & & \\ 
     PliableBVS  & 0.001 (0.000) &0.970 (0.068) & 1.000 (0.000) & 0.999 (0.001)& 0.000 (0.000)& 10 (7) &0.313 (0.312)  \\ 
     PliableLasso & 0.004 (0.001) & 0.941 (0.087)& 0.962 (0.018) &  0.962 (0.018)& 0.038 (0.018)& 91 (19) &2.277 (0.534)  \\ 
     Scenario 5 & $p=20$, $SNR=2$&  & & & &  &\\ 
     PliableBVS  & 0.034 (0.010) &0.910 (0.082) & 0.959 (0.030) & 0.950 (0.029)& 0.041 (0.029)& 12 (7) &1.047 (0.466)  \\ 
     PliableLasso & 0.050 (0.011) & 0.993 (0.021)& 0.566 (0.130) &  0.643 (0.106)& 0.433 (0.129)& 17 (36)&1.129 (0.325)  \\ \hline
    \end{tabular}
    }
    \begin{tablenotes}
    \item[1] $\mathbf{B}$ represents the coefficient vector of the main and interaction coefficients together. 
    \item[2] Sensitivity is the proportion of non-zero coefficients correctly estimated as non-zeros. 
    \item[3] Specificity is the proportion of zero-coefficients correctly estimated as zeros.
    \item[4] Accuracy denotes the percentage of both true non-zeros that are correctly\\ estimated as non-zeros and true zeros correctly estimated as zeros. 
    \item[5] \# Non-zeros represents the average number of selected features.
The number of interaction features is\\ shown in parentheses.
For PliableBVS, feature selection was based on the MPM from each repetition.\\ Reported features are those selected in at least 6 of the 50 repetitions. 
    \item[6] The MSE on an independent test data set simulated according to the same generative model as the \\training data sets.
    \item[*] All values are reported as means (standard deviations) across 50 simulation repetitions.

  \end{tablenotes}
    \end{threeparttable}
  \end{center}
\end{table}

\subsection{Simulation under logistic regression }
We generated the data using the same data-generating mechanisms  scenarios 1-3 as in the linear regression case. We also presented a fourth example where we followed scenario 1 but with $p=500$ to represent a high-dimension case and with a signal-to-noise ratio of 2.
Each generated continuous response was transformed into a binary outcome by applying the logistic function to the linear predictor and assigning the outcome as 1 when the resulting probability exceeded 0.5, and 0 otherwise. Each scenario was repeated 50 times. All analysis were carried out using the logistic regression of the pliable lasso and the logistic regression option of the proposed PliableBVS.  

The summary of the logistic regression simulations is shown in Table \ref{table31}. The results closely mirror those observed in the linear regression setting, with PliableBVS demonstrating superior performance in both predictive accuracy and variable selection, particularly in moderate to high SNR settings. The results of the simulation experiments can be summarized as follows:
\begin{itemize}
    
    \item In all scenarios across the range of SNR settings, PliableBVS performs better in terms of the binomial deviance based on the test data set. For outcome classification in terms of the area under the receiver operating characteristic curve (AUC), PliableBVS also outperforms PliableLasso. 
   
    \item PliableBVS also outperforms PliableLasso in estimating the true values of both main effects and interaction effects in terms of the bias of estimated coefficients $(1/(p+p\times K))\lVert \hat{\mathbf{B}}-\mathbf{B}\rVert_1$ in table \ref{table31}.
    \item For variable selection in terms of sensitivity and specificity, it can be seen that PliableBVS is generally able to identify both the zeros and non-zeros better than the PliableLasso. 
    
\end{itemize}

\begin{table}[ht]
\small 
\centering
  \begin{center}
    \caption{Results from the logistic regression simulation examples. }
    \label{table31}
    \begin{threeparttable}
    \resizebox{\textwidth}{!}{%
    \begin{tabular}{|l l l l l l   l|} 
    \hline
     \textbf{Model} & $(1/(p+p\times K))\lVert \hat{\mathbf{B}}-\mathbf{B}\rVert_1$\tnote{1*} & Sensitivity\tnote{2*} & Specificity\tnote{3*} & AUC\tnote{4*}  &\# Non-zeros\tnote{5} & Deviance \tnote{6*} \\
      \hline
      Scenario 1 & $p=20$ , $SNR=2$ & & & &  &\\
     PliableBVS & 0.111 (0.013) & 0.950 (0.113) & 0.925 (0.047) &  0.920 (0.047) & 6 (7) &3.685 (1.135) \\
     PliableLasso  & 0.125 (0.009)& 0.977 (0.121) & 0.806 (0.159) & 0.903 (0.064) & 11 (13) &4.192 (1.312) \\
     Scenario 2 & $p=20$ , $SNR=2$ & & &  & &\\
     PliableBVS  &0.121 (0.010) & 0.750 (0.144) & 0.914 (0.053) &  0.920 (0.022)  & 5 (7) &3.851 (0.617)\\
     PliableLasso & 0.097 (0.013) &0.653 (0.247) & 0.895 (0.174) & 0.900 (0.053)  & 7 (6) &4.542 (1.603)  \\
     Scenario 3  & $p=20$ , $SNR=6$ & &  & & &\\ 
     PliableBVS  & 0.061 (0.013) & 0.924 (0.069) &0.856 ( 0.079) &  0.933 (0.016)  & 11 (18) &3.409 (0.471)  \\ 
     PliableLasso   & 0.069 (0.013) & 0.874 (0.178) & 0.743 (0.184) & 0.913 (0.033)  & 14 (22) &3.925 (0.827)   \\ 
     Scenario 1 &$p=500$ , $SNR=2$ & &  & &  &\\ 
     PliableBVS  & 0.005 (0.000) &0.817 (0.162) & 0.999 (0.001) & 0.879 (0.077)& 4 (2) &4.838 (1.716)  \\ 
     PliableLasso & 0.006 (0.001) & 0.747 (0.216)& 0.991 (0.016) &   0.776 (0.095)& 25 (3) &6.868 (1.715)  \\ 
        \hline
    \end{tabular}
    }
    \begin{tablenotes}
    \item[1] $\mathbf{B}$ represents the coefficient vector of the main and interaction coefficients together. 
    \item[2] Sensitivity is the proportion of non-zero coefficients correctly estimated as non-zeros. 
    \item[3] Specificity is the proportion of zero-coefficients correctly estimated as zeros.
    \item[4] Area under the curve (AUC) from the receiver operating characteristic (ROC) curve. 
    \item[5] \# Non-zeros represents the average number of selected features.
The number of interaction features is\\ shown in parentheses.
For PliableBVS, feature selection was based on the MPM from each repetition.\\ Reported features are those selected in at least 6 of the 50 repetitions. 
    \item[6] The binomial deviance on an independent test data set simulated according to the same generative model \\as the training data sets.
\item[*] All values are reported as means (standard deviations) across 50 simulation repetitions.
  \end{tablenotes}
    \end{threeparttable}
  \end{center}
\end{table}

\section{Real data analysis}\label{real}

In this section, we apply the two variable selection methods, Bayesian pliable‑lasso (PliableBVS) and the frequentist pliable lasso (PliableLasso), to two high-dimensional biomedical studies. 
The labor onset data set from \cite{labor} has continuous outcomes and is modeled by a linear regression, 
and the late-onset preeclampsia (LOPE) data set \citep{Degnes2024} has binary outcomes and is modeled by a logistic regression.

\subsection{Labor onset data}
The purpose of this study was to predict the time to spontaneous labor onset (excluding those with medically induced labor and those who  underwent cesarean section without labor) using data from 63 women and measurements taken during their last 100 days of pregnancy. Proteomics and metabolomics data were obtained from plasma samples. 
The proteomics data contains measurements for 1322 proteins and the metabolomics data for 3529 metabolites. 
The assumption is that some metabolomics or proteomics  variables can directly be associated with pregnancy length and thus time to spontaneous labor. Hence, identifying such potential biomarkers  can help to improve understanding of the variations in pregnancy length and improve prediction of expected time to spontaneous labor onset. 

We analyzed this data set under three possible scenarios. 
(i) We considered $X$ as metabolomics and included the different measurement time points as $Z$. Plasma samples were collected at three time points measured relative to labor onset: T1 (mean = 77 days before onset, SD = 20 days), T2 (mean = 33 days before onset, SD = 22 days), and T3 (mean = 15 days before onset, SD = 11 days). Thus, later time points correspond to shorter times to labor onset. It is therefore reasonable to expect that the effects of some features may vary substantially over time. We included the time points as categorical variables by encoding them as two dummy variables, representing comparisons of T2 versus T1 and T3 versus T1, respectively.
(ii) Proteomics features were used as $X$ and time points as $Z$. 
(iii) Exploring potential interaction effects between proteomics and metabolomics data. For this scenario, we only used the data from the first time point, and considered $X$ as proteomics data and $Z$ as a few selected metabolites, which were identified as associated with labor onset in \cite{labor}. 
The analysis for each scenario was repeated 10 times, each involving a different random splitting of the data into training and testing, thus reducing the dependence of the results on the choice of the training-test split and providing an assessment of the sensitivity of the results to this choice, in particular with respect to the stability of the variable selection. 
The results are shown in table \ref{table4}. In all three scenarios, PliableBVS performs better than PliableLasso in terms of test MSE. PliableBVS also selected fewer main and interaction features.

\begin{figure}[h!]

  \centering
  \includegraphics[width=0.99\textwidth]{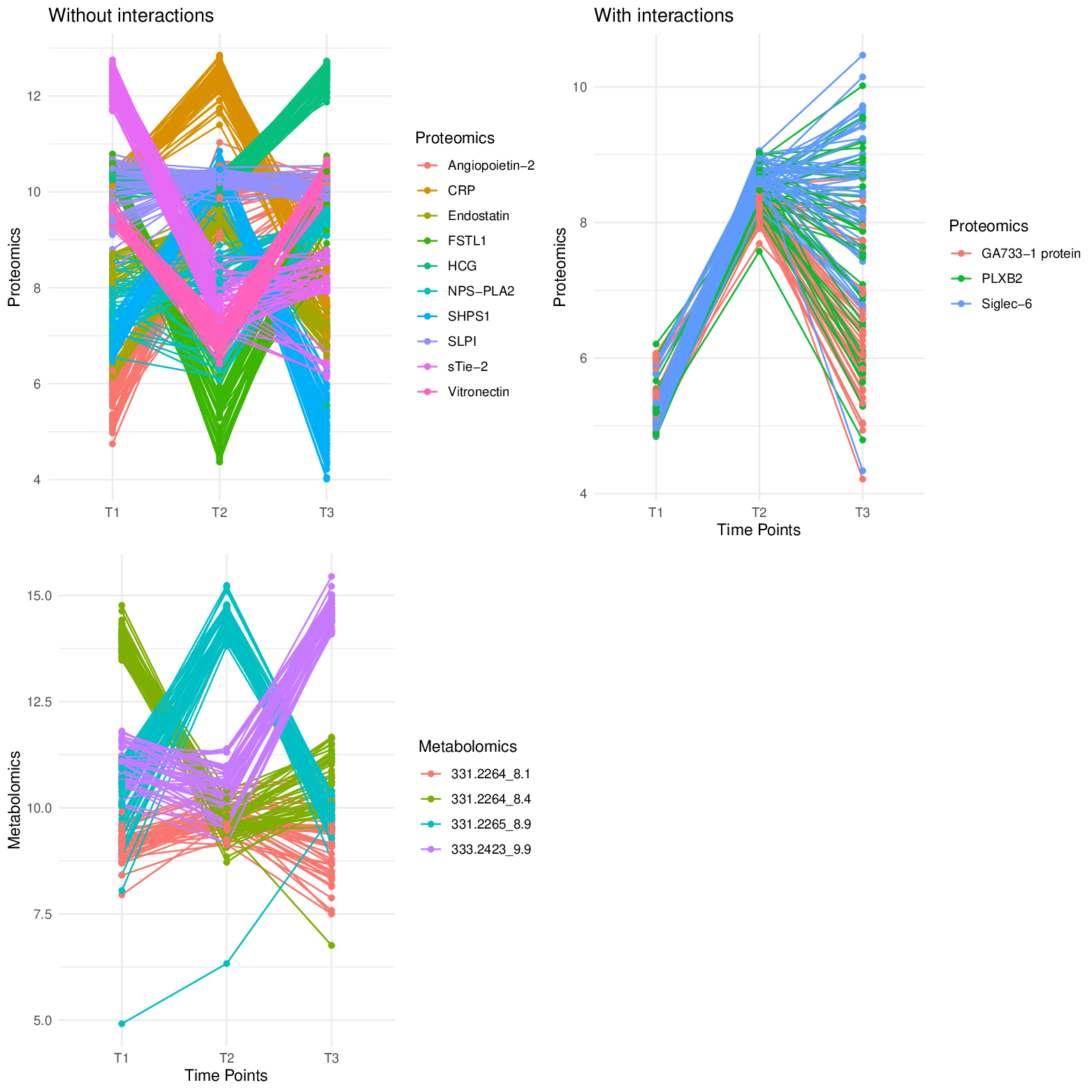}


\caption{Results for the analysis of the labor onset data set \cite{labor}: trajectories of the measurements over time of those maternal metabolomics (bottom row, scenario 1) and proteomics (top row scenario 2) features that were identified by the PliableBVS models. Features identified with main effect without interactions (left) and and with non-zero interactions (right). }\label{features_time}
\end{figure}

We investigated the features of the PliableBVS model for potential biomarkers to predict the time to spontaneous onset of labor. In scenario (i), PliableBVS included the metabolomics features at different times but did not include any interaction term with time, indicating that the metabolite effects on labor onset remain stable over time. The figure on the bottom left of figure \ref{features_time} shows the metabolites with identified non-zero main effects for all patients over the three time points. Indeed, the observed metabolite values remain rather similar for all pregnancies at each of the three time points, indicating no significant change in effect size that might indicate an interaction with time. 

\begin{figure}[h!]

  \centering
  \includegraphics[width=0.99\textwidth]{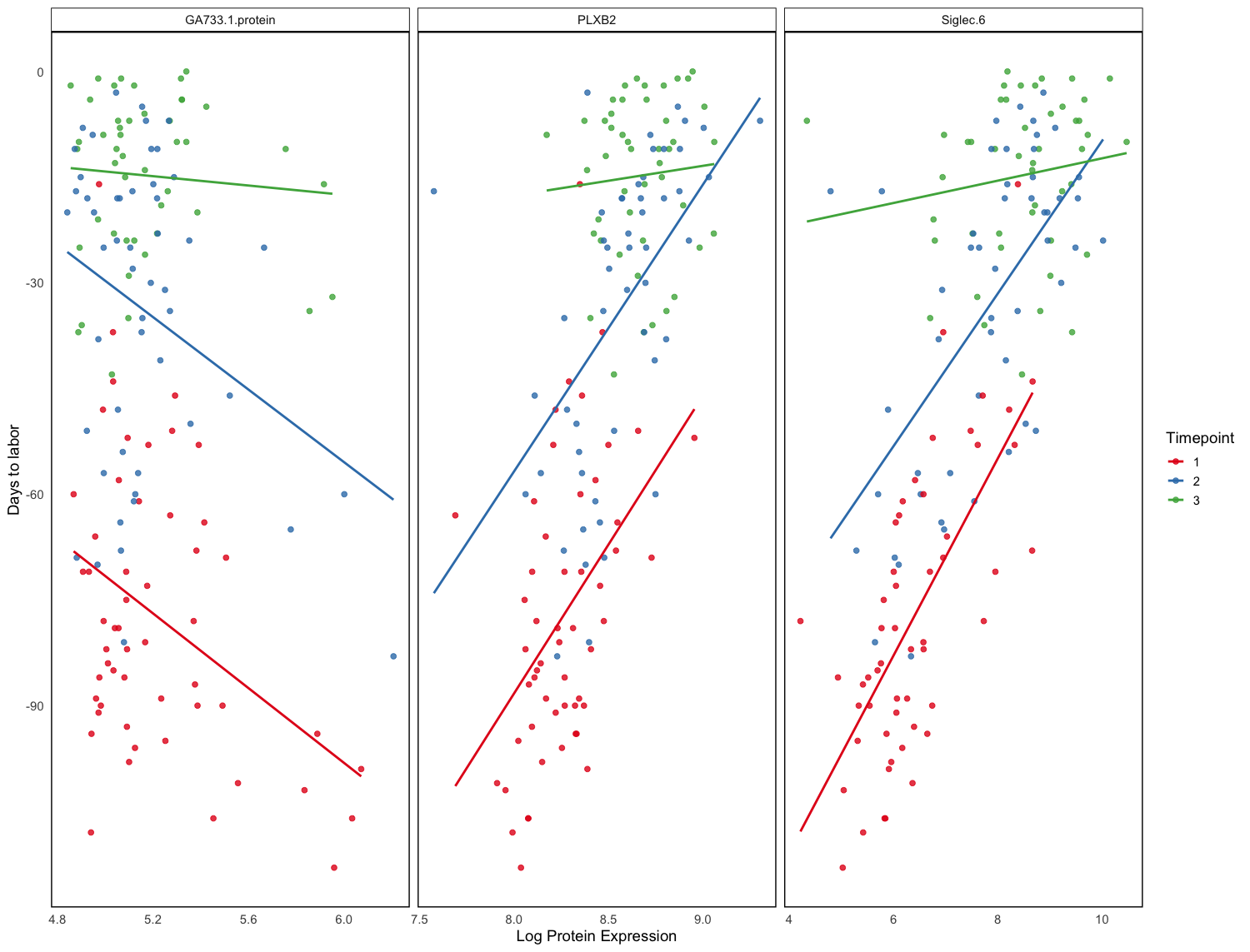}


\caption{Results for the analysis of the labor onset data set \cite{labor} from scenario (ii):  Scatter plots of three maternal  proteomics features that were identified by the PliableBVS models that allowed for interactions with time and the days to labor. univariate linear regression lines are added to show the relationship between the protein variables and the response at each of the time points. For all variables, the relationship seems to be most strongly either positive or negative at time points 1 and 2, and reduced for the time point 3.}\label{scater_time}
\end{figure}

In scenario (ii), we see a similar pattern among the identified proteins without interaction effect over time in the figure on the top left. Again, the explanation could be that their effects remain similar over time as explained by \cite{labor} or  there is possibility of higher-order or time-specific interaction effects, which may warrant further investigation.
A different pattern is seen in the figure on the top right, which represents the proteins for which non-zero interaction effects with time are found. While the observed protein abundance levels are rather similar for all pregnancies at the first two time points, they diverge strongly for time point three, which indicates a change in the strength of association at time point 3 between the abundance measured for these proteins and time to labor onset. This can be clearly seen in figure \ref{scater_time}. 
Among the proteins identified by pliableBVS, when time point was included as a modifying variable, and also reported by \cite{labor}, are Angiopoietin-2, Endostatin, SLPI, Sialic acid-binding immunoglobulin-like lectin–6 (Siglec-6), soluble Tie2 receptor (sTie-2), and Plexin-B2 (PLXNB2). These proteins have previously been found in other related studies \citep{coop1,labor2,labor3}, showing their potential contribution to the prediction of time to spontaneous labor onset. Other features identified by our model, but not found in the work of \cite{labor}, are C-reactive protein (CRP), which levels has been found to increase significantly as labor approached \citep{CRP1}; Secretory Leukocyte Protease Inhibitor (SLPI) \citep{labor6,labor7}; and vitronectin \cite{labor8}. The results implies that the proposed model can help to identify relevant features. 
\begin{table}[ht]
\small 
\centering
  \begin{center}
    \caption{Results for the analysis of the labor onset data: summary of prediction and variable selection performance measures across 10 random splits into training and test data. }
    \label{table4}
    \begin{threeparttable}
    \resizebox{\textwidth}{!}{%
    \begin{tabular}{|  l l l l l|} 
    \hline
     \textbf{Model}  & \multicolumn{2}{|c|}{Test MSE\tnote{1}}   & \multicolumn{2}{c|}{\# Non-zeros\tnote{2}}  \\
      &\multicolumn{1}{|c} {Mean} & \multicolumn{1}{c|}{SD} &\multicolumn{1}{|c} {Main effect} & \multicolumn{1}{c|}{Interaction effect} 
      \\

      \hline
      (i) $X$ as metabolomics, $Z$ as time &   & & & \\
     PliableBVS  & 374.579  &  165.659 &  4  &0 \\
     PliableLasso  & 687.851  & 131.198 & 18   &0 \\
  
    (ii) $X$ as proteomics, $Z$ as time&  & & & \\
     PliableBVS   & 377.474  & 162.702 &  13  &3 \\
     PliableLasso  &493.130  &   223.389 & 78   &12   \\

     (iii) $X$ as proteomics, $Z$ as metabolomics &  & & & \\
     PliableBVS   & 278.412  & 106.176 &  10  &2 \\
     PliableLasso  &353.948  &  242.013 & 28   &5   \\
        \hline
    \end{tabular}
    }
    \begin{tablenotes}
    \item[1] The MSE on an independent test data set.
    \item[2] Non-zeros represents the number of features selected. For PliableBVS, this was obtained using the MPM  results from each repetition. 
    
  \end{tablenotes}
    \end{threeparttable}
  \end{center}
\end{table}

\subsection{Preeclampsia data}
This was a study conducted to identify new proteomic biomarkers associated with late-onset preeclampsia (LOPE) \citep{Degnes2024}. This is a longitudinal study, covering samples taken during three different visits at gestational age in weeks 12-19, 20-26, and 28-34 for 35 women with LOPE (diagnosis $\geq 34$ gestational weeks) and 70 healthy pregnant women. The data set comprises 4565 proteins, clinical variables that include maternal age, pre-pregnancy BMI, history of smoking (yes or no), nulliparity (parity zero or not) and sex of the baby (female or not). The binary response indicates whether a woman developed LOPE. 

We considered two modeling scenarios.

(i) Proteins (X) were treated as potential main effects, while clinical variables (Z) were included as potential modifying variables. The inclusion of clinical variables as mandatory terms was previously considered by \cite{Degnes2024}. In this analysis, only data from the third visit were used.

(ii) In the second scenario, data from all three visits were incorporated into the model. The visit variable was included as a categorical modifying variable and encoded using two dummy variables corresponding to the comparisons of V2 versus V1 and V3 versus V1, respectively. This formulation allows protein effects to vary across gestational age and facilitates the identification of proteins whose associations with LOPE change over time.

\begin{figure}[ht]
\begin{subfigure}{.55\textwidth}
  \centering
   \includegraphics[width=0.9\textwidth]{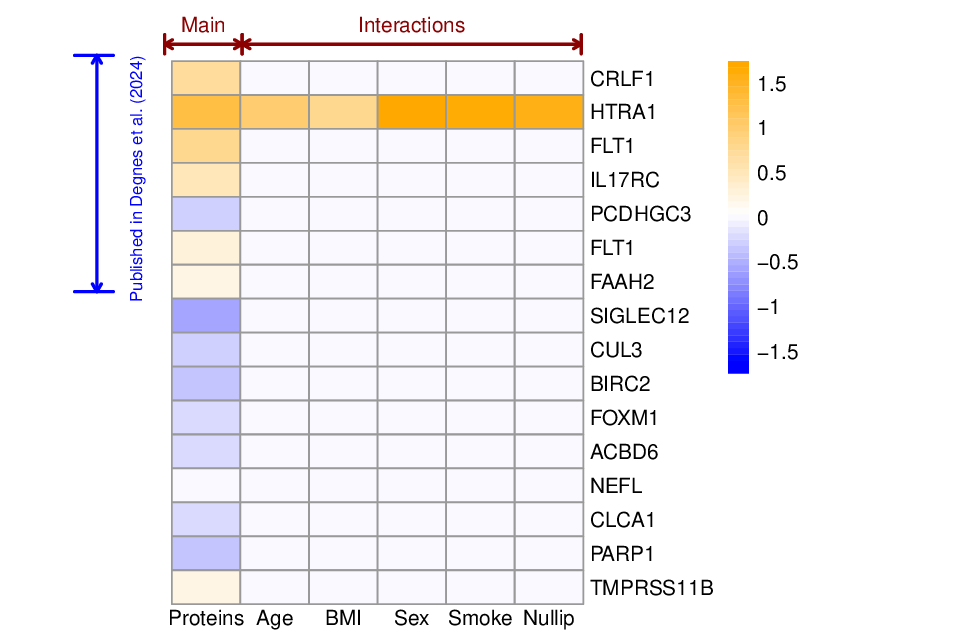}\caption{}\label{interplot3}
\end{subfigure}
\begin{subfigure}{.55\textwidth}
  \centering
  \includegraphics[width=0.9\textwidth]{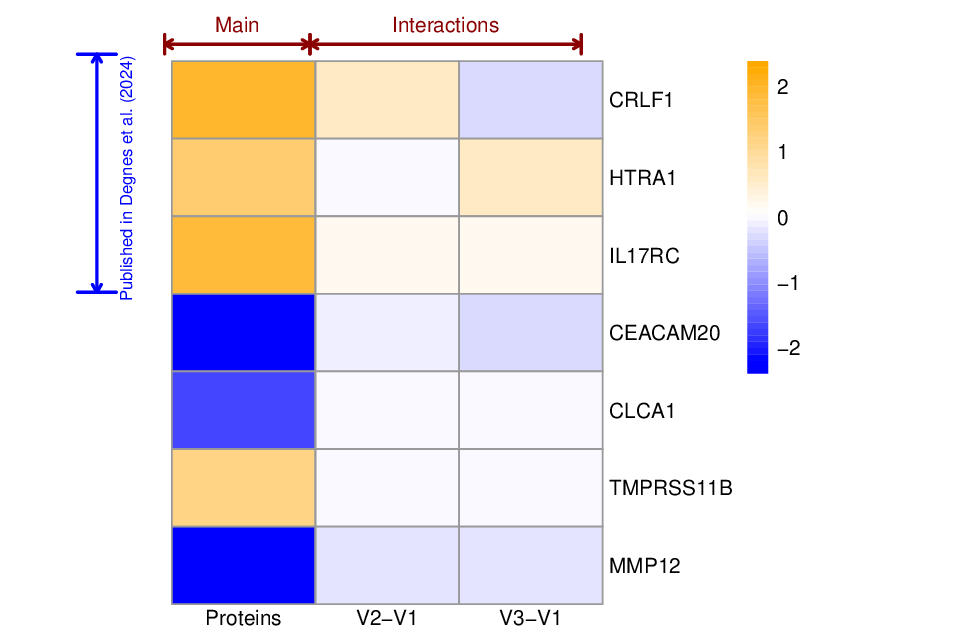}
 \caption{}\label{interplot2}
\end{subfigure}
\caption{Proteins' estimated main effects for classifying LOPE and their interaction effects with (a) clinical data and (b) time. We have indicated the common proteins found in both our model and form the work of \cite{Degnes2024}. In (a), the first FLT1 is FLT1.16315.105 and the second is FLT1.8231.122. In (b),  Visit 1 (V1) represents gestational age in weeks 12-19, visit 2 (V2) 20–26 and visit 3 (V3) 28–34. }
\end{figure}

Figure \ref{interplot3} summarizes the results of the analysis for modeling scenario (i). The following proteins were selected by our PliableBVS model and were also found in the work of \cite{Degnes2024}: cytokine receptor-like factor 1 (CRLF1), Serine protease HTRA1, Soluble Fms-like tyrosine kinase 1 (FLT1), Interleukin-17 Receptor C (IL17RC), Protocadherin Gamma Subfamily C 3 (PCDHGC3) and Fatty Acid Amide Hydrolase 2 (FAAH2). Although HTRA1, FLT1 and IL17RC have also been reported to be associated with PE elsewhere \citep{GESUITA201958,ijms20174246,elshahaway2019role}, FAAH2 was previously reported as a possible biomarker for LOPE only by \cite{Degnes2024}. The findings of these proteins confirm the model's ability to identify relevant biomarkers. In addition to these proteins, the model also identified other proteins that were not identified by \cite{Degnes2024} (Figure \ref{interplot3}). Among these are Forkhead Box M1 (FOXM1), which showed a negative association with LOPE. Studies have consistently reported that FOXM1 mRNA and protein levels are significantly reduced in placental tissues in preeclamptic pregnancies compared to those in healthy pregnancies \citep{Cui2018FOXM1Preeclampsia}. This could be related to its role in trophoblast proliferation and invasion. In terms of clinical risk factors, maternal age showed a negative association with LOPE as also shown in \cite{Degnes2024}, indicating that younger women have higher risk of developing LOPE. However, the relationship between maternal age and preeclampsia has been described as a spoon-shape, meaning that both young and older women have an increased risk of preeclampsia \citep{https://doi.org/10.1002/uog.12494}. One possible explanation for the observed negative association is that preeclampsia occurs more frequently among nulliparous women, who also tend to be younger on average. in addition, this discrepancy could be due to the sub-sampling of participants in the STORK cohort used in \cite{Degnes2024}, where the maternal ages were lower in the LOPE group than in the healthy group (the maternal age ranged between 24-41 (mean = 32.1) in the healthy group and 23-39 (mean 29.8) in LOPE group with a statistically significant lower mean maternal age in LOPE compared to the healthy group (p = 0.01). Therefore, the validity of our results may be limited to the observed age range. Moreover, BMI and nulliparity have positive associations with LOPE, which corresponds to the results of the work of \citep{Degnes2024} and also to the literature \citep{Morikawa2013,VATS2021536,JUNG2022S844}. Another interesting finding from our model is the negative association of smoking status with LOPE. Many previous studies have also shown that smoking is associated with a lower incidence of preeclampsia, particularly late-onset \citep{conde2008smoking}. Smoking has also been associated with reduced circulating levels of sFlt-1 and sEng, biomarkers implicated in the development of preeclampsia \citep{jeyabalan2008cigarette,doi:10.1161/HYPERTENSIONAHA.109.148973}. However, these findings should not be interpreted as evidence of a protective clinical recommendation for smoking, given the well-established adverse maternal and fetal health consequences associated with tobacco exposure during pregnancy. Note that we identified interactions (modifying effects) of these clinical variables with only one protein: we observed positive interactions between HTRA1 and all clinical variables in the model (Figure \ref{interplot3}). These are positive interaction effects, suggesting that the association between HTRA1 and LOPE varies positively across the levels or values of the corresponding clinical variables.

\begin{figure}[h!]

  \centering
  
 \includegraphics[height=10cm,width=16cm]{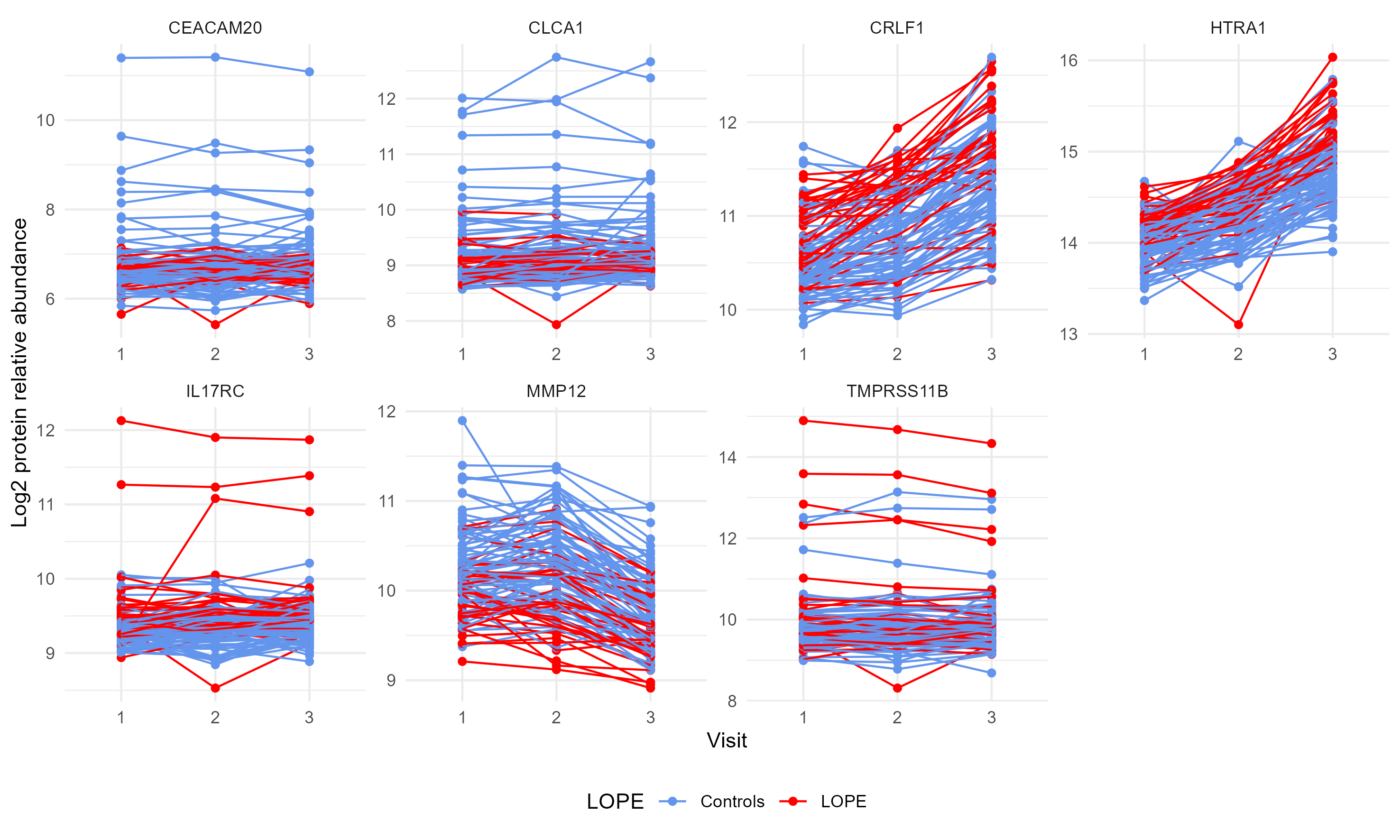}

\caption{Preeclampsia study \citep{Degnes2024} scenario (ii): Longitudinal protein expression profiles in control and late-onset preeclampsia (LOPE) groups at three visits across gestation (gestional age in weeks 12-19, 20-26 and 28-34, respectively). Line plots show the $\log_2$-transformed relative abundance of selected proteins across three visits per individual participants. Blue color indicate protein trajectories in healthy control pregnancies, while red color indicate trajectories in pregnancies that would develop late-onset preeclampsia after V3. }\label{interplot}
\end{figure}

Figures \ref{interplot2} and \ref{interplot} display the results for the modeling scenario (ii).  
Our PliableBVS model identified CRLF1, HTRA1, IL17RC as proteins associated with LOPE. All three proteins showed positive regression coefficients in predicting LOPE. These proteins were also identified by \cite{Degnes2024}. Regarding a potential change in their contributions to the predicted LOPE risk across visits, our analysis of the interactions with visits revealed a gestational age–dependent shift in the association between CRLF1 abundance and the probability of developing LOPE (figure \ref{interplot2}). During visit 2 (gestational weeks 20–26), CRLF1 has a positive change in the regression coefficient (positive interaction effect), indicating that higher abundance is associated with increased risk of LOPE. However, by visit 3 (weeks 28–34), the coefficient change becomes negative (negative interaction effect), suggesting that higher CRLF1 abundance later in pregnancy may be associated with a reduced risk of LOPE. We speculate whether this shift may reflect a dynamic biological role, where CRLF1 contributes to early pathological processes but later participates in adaptive or compensatory mechanisms that mitigate disease progression. The protein HTRA1 showed a positive main effect (positive effect on the risk of LOPE), but had no significant interaction effect with gestational age during visit 2 indicating a consistent risk pattern. However, during visit 3, HTRA1 also showed a positive interaction with gestational age, indicating that its influence on LOPE risk increases as pregnancy progresses. The effect of IL17RC on the prediction of LOPE increased at visit 3. These results support the findings of \cite{Degnes2024} who analyzed data from the three visits separately. Other proteins identified by the model include CEACAM20, CLCA1, TMPRSS11B, and MMP12. Among these, CLCA1 and TMPRSS11B did not exhibit interaction effects with visit. Compared with earlier analyses of the same dataset, these proteins may have been identified due to the proposed model’s ability to simultaneously capture both main and interaction effects under a hierarchical variable selection framework instead of modeling each time point separately.

CLCA1 showed a negative association with LOPE, indicating that higher expression levels were associated with lower observed LOPE prevalence in our data. CLCA1 belongs to the calcium-activated chloride channel family and is primarily expressed in epithelial tissues, where it has been implicated in mucus production, ion transport, and immune regulation \citep{CLCA1}.

CEACAM20 also exhibited negative main and interaction effects, with increasing interaction effects from visit 2 to visit 3. Although CEACAM20 has not previously been studied in the context of preeclampsia, our findings suggest that it may represent a potential biomarker associated with LOPE-related pathways. This observation is broadly consistent with previous studies implicating CEACAM1 in pregnancy regulation \citep{https://doi.org/10.1111/aji.13375}.

Matrix Metallopeptidase 12 (MMP12) similarly exhibited negative main effects, with inverse associations remaining relatively consistent across visits 2 and 3. MMP12 also showed negative interaction effects with gestational age. There is evidence on associations between MMPs and preeclampsia, although very scarce on MMP 12 in particular \citep{YAKOVLEVA202236}.

Overall, these findings should be interpreted cautiously and regarded as hypothesis-generating. Further experimental and clinical studies are required to validate the identified proteins and clarify their potential biological roles in LOPE. Figure \ref{interplot} shows the longitudinal protein expression profiles in controls and the LOPE group across visits. Group-level trends suggest differential expression dynamics across gestation between the two groups, particularly for CRLF1 and HTRA1, which reflects the strong time interaction effects for these two proteins.

\clearpage
\section{Conclusion}\label{conclusion}

In this article, we introduced the Bayesian pliable lasso model for variable selection (PliableBVS) of high-dimensional covariates, which allows the inclusion and detection of interactions between selected high-dimensional covariates and low-dimensional mandatory modifying variables. 
Building upon the pliable lasso framework, we employed structured Bayesian priors that combine the Bayesian lasso (Laplace prior) with a spike-and-slab prior for both main effects and interaction terms, thereby enforcing an asymmetric weak hierarchical structure while achieving efficient variable selection. 
In addition to developing the model for continuous outcomes, we extended it for binary outcomes by means of the P\'olya-gamma augmentation for logistic regression. 

Our simulation studies for both continuous and binary outcomes demonstrated that PliableBVS provides accurate variable selection with lower false discovery rates than the frequentist pliable lasso across all simulation scenarios. 
In particular, PliableBVS was able to identify truly relevant variables in both high and low signal-to-noise ratio was, confirming its robustness in realistic settings. 

The two applied examples from molecular medicine, focusing on the identification of omics signatures predictive of time to spontaneous labor onset (proteomics and metabolomics) and late-onset preeclampsia (proteomics), demonstrated that PliableBVS is capable of identifying biologically relevant main effects and interactions. However, as with other high-dimensional discovery analyses in pregnancy cohorts, these findings should be regarded as exploratory and cohort-specific, requiring validation in independent populations before clinical relevance can be established. 

As with other lasso-type models, PliableBVS tends to select one variable from a set of highly correlated predictors.
A possible extension would be to replace the Laplace prior with an elastic-net-type prior \citep{Li2010}, which blends Gaussian and Laplacian components and can select correlated variables simultaneously. Correspondingly, one could introduce selection of hyper priors for $\rho$ and $\zeta$ that encourages the joint selection of features, for example by replacing the beta priors with Markov random field (MRF) prior \citep{Zhao2021}.
It is also straightforward to adapt the proposed models to simple weak hierarchical interaction structure by removing the conditioning on the selection of high-dimensional covariates, since the low-dimensional mandatory modifiers are always included. 

The PliableBVS package is available on Github (https://github.com/Theo-qua/PliableBVS) with version 0.1.0 used throughout this paper, and a web-based API (https://bpliable-api.onrender.com/\_\_docs\_\_) is provided for non-R users.
Overall, the Bayesian pliable lasso offers a flexiable, interpretable, and computationally efficient tool for high-dimensional regression with structured interactions, opening new possibilities for personalized medicine and other domains where heterogeneous effects are modulated by a few mandatory variables.

\section*{Acknowledgements}
This work received funding from the European Union’s Horizon 2020 Research and Innovation program, under the Marie Skłodowska-Curie Actions Grant, agreement No. 801133 (Scientia fellowship), and under grant agreement No. 847912 (``RESCUER"). It has also received funding from the Innovative Medicines Initiative 2 Joint Undertaking (JU) under grant agreement No 853988 (``ìmSAVAR"). The JU receives support from the European Union’s Horizon 2020 research and innovation programme and EFPIA and JDRF INTERNATIONAL.

\bibliographystyle{apalike} 
\bibliography{refs}

\FloatBarrier 
\section{Supplementary Material}
\beginsupplement
\subsection{Gibbs sampler fro the logistic regression}\label{suplement}
\begin{itemize}

\item {\bf Step 1 (Gibbs for main effects $\beta$ and variable selection):}\\  Let $\beta_{-j}$ be coefficients of the main effects without the $\text{j}^{\text{th}}$ component, i.e. \\ $\beta_{-j}=(\beta_1,\beta_2,\cdots,\beta_{j-1},\beta_{j+1},\cdots, \beta_p)$ and $X_{-j}$  be the covariate matrix without the $\text{j}^{\text{th}}$ column. Also let $\bar{O}=O-\beta_0\mathbf{1}-Z\boldsymbol{\theta}_0-\sum_j(X_j \circ Z)\boldsymbol{\theta}_j$ and $\mu_j=S_jX_j^T\Omega(\bar{O}-X_{-j}\beta_{-j})$, $S_j=(X_j^T\Omega X_j+\frac{1}{\sigma^2\tau_{\beta_j}^2})^{-1}$. Then the conditional posterior distribution of $\beta_j$ is a spike-and-slab distribution 

\begin{equation}\label{batejupdate2}
    \beta_j|\ldots\sim (1-\eta_j)\mathcal{N}(\mu_j,S_j)+\eta_j\delta_0(\beta_j), \ \ j=1,\cdots,p, 
\end{equation}
 where
\begin{equation}\label{l_j_update2}
    \eta_j=p(\beta_j=0|\ldots)=\frac{\rho}{\rho+(1-\rho)(\tau_{\beta_j}^2\sigma^2)^{-\frac{1}{2} } S_j^{\frac{1}{2}} \text{exp}\biggl\{ \frac{1}{2}  S_j \left( X_j^T\Omega (\bar{O}-X_{-j}\beta_{-j}) \right)^2 \biggr\} }. 
\end{equation}

\item {\bf Step 2 (Gibbs for the interaction effects $\theta$ and variable selection):}\\ For the $\theta_{jk}$ update we define $\bar{O}$ as $\bar{O}=O-\beta_0\mathbf{1}-Z\boldsymbol{\theta}_0-X\boldsymbol{\beta}$ and let $\mu_{jk}=S_{jk}(X_j \circ Z_k)^T\Omega(\bar{O}-\sum_{w\neq k}(X_j \circ Z_w)\theta_{jw})$, $S_{jk}=((X_j \circ Z_k)^T\Omega(X_j \circ Z_k)+\frac{1}{\sigma^2\tau_{\theta_{jk}}^2})^{-1}$. Then the conditional posterior distribution of $\theta_{jk}$ given that $\beta_j\neq 0$ is a spike-and-slab distribution 

\begin{equation}\label{theta_j_upate}
    \theta_{jk}|\ldots \sim (1-\gamma_{jk})\mathcal{N}(\mu_{jk},S_{jk})+\gamma_{jk}\delta_0(\theta_{jk}),
\end{equation}
 where
\begin{equation}\label{r_jk_update}
    \gamma_{jk}=p(\theta_{jk}=0|\ldots)=\frac{\zeta}{\zeta+(1-\zeta)(\tau_{\theta_{jk}}^2 \sigma^2)^{-\frac{1}{2} } S_{jk}^{\frac{1}{2}} \text{exp}\biggl\{ \frac{1}{2}   S_{jk}\left( (X_j \circ Z_k)^T\Omega (\bar{O}-\sum_{w\neq k}(X_{j} \circ Z_w)\theta_{jw}) \right)^2 \biggr\} }. 
\end{equation}

\item {\bf Step 3 (Gibbs for $\sigma^2$ ):}\\  The  conditional posterior distribution of  $\sigma^2$ is conditionally inverse gamma, i.e., 

\begin{multline}\label{sigma}
    \sigma^2 | \ldots \sim \text{Inverse Gamma} \biggr(  \frac{\sum_jQ_j}{2} +\frac{\sum_{jk}R_{jk}}{2} +\lambda_1, \\ 
    \frac{1}{2}\biggl[  \boldsymbol{\beta}^T D_{\tau_\beta}^{-1} \boldsymbol{\beta} +\sum_j\boldsymbol{\theta}_j^T D_{\tau_{\theta_j} }^{-1}\boldsymbol{\theta}_j    \biggr]+\lambda_2  \biggl).
\end{multline}

\item {\bf Step 4 (Gibbs for $\Omega$ ):}\\  For  $\Omega=(\omega_1, \dots, \omega_N)$ the  conditional posterior distribution of $\omega_i$ also follows the P\'olya-gamma distribution as 

\begin{equation}\label{omegapost}
    \omega_i | \ldots \sim \text{PG} \biggr( 1, \beta_0+Z_i^T\boldsymbol{\theta}_0+X_i^T\boldsymbol{\beta}+\sum_{j=1}^pX_{ij}  Z_i^T\boldsymbol{\theta}_j \biggl).
\end{equation}

\item {\bf Step 5 (Gibbs for $\beta_0$ ):}\\  Let $\bar{O}=O-Z\boldsymbol{\theta}_0-X\boldsymbol{\beta}-\sum_j(X_j \circ Z)\boldsymbol{\theta}_j$. The conditional posterior distribution of $\beta_0$ follows the normal distribution as 
\begin{equation}\label{beta0post2}
    \beta_0 | \ldots \sim \mathcal{N} \biggl( \frac{\sum_{i=1}^N\bar{O}_i\omega_i}{\sum_{i=1}^N\omega_i+\frac{1}{c^2}}, \biggl(\sum_{i=1}^N \omega_i+\frac{1}{c^2} \biggr)^{-1} \biggr).
\end{equation}


\item {\bf Step 6 (Gibbs for $\boldsymbol{\theta}_0$ ):}\\  Let $\bar{O}=O-\beta_0\mathbf{1}-X\boldsymbol{\beta}-\sum_j(X_j \circ Z)\boldsymbol{\theta}_j$. The conditional posterior distribution of $\boldsymbol{\theta}_0$ follows the normal distribution as 
\begin{equation}\label{theta0post2}
    \boldsymbol{\theta}_0 | \ldots \sim \mathcal{N} \biggl( 
    \biggl[Z^T\Omega Z+\frac{1}{v^2}\mathbb I_K\biggr]^{-1} Z^T\Omega\bar{O}, 
    \biggl[Z^T\Omega Z+\frac{1}{v^2}\mathbb I_K\biggr]^{-1} \biggr).
\end{equation}
\end{itemize}

\end{document}